\documentclass[aps,prl,reprint,groupedaddress]{revtex4-2}

\usepackage{graphicx}
\usepackage{epstopdf}
\usepackage{amsmath}
\usepackage{bm}
\usepackage{epsf,epsfig,graphics}
\usepackage{float}
\usepackage{makeidx}
\usepackage{latexsym}
\usepackage{amssymb}
\usepackage{subfigure}
\usepackage{amsmath}
\usepackage{amsfonts}
\usepackage{graphicx}
\usepackage{dcolumn}
\usepackage{amsmath}
\usepackage{color}
\usepackage{subfigure}
\usepackage{natbib}
\usepackage{booktabs}
\usepackage{multirow}
\usepackage{bm}% bold math
\usepackage{makecell}
\usepackage{xcolor}

\begin{document}

\title{Dynamics of Information Exchange in Zebrafish: The Role of U-Turns in Visual Communication and Behavior Modulation }

\author{C. K. Chan\footnote{ckchan@gate.sinica.edu.tw}}
\author{Hao-Yun Hsu }

\affiliation{Institute of Physics, Academia Sinica, Nankang, Taipei 115, Taiwan, R.O.C.}


\date{\today}
\begin{abstract}
Motions of visually coupled zebrafish pairs are studied to understand the effects of visual information exchange on their behavior as a function of their minimal separation ($d$). We find that when $d$ is small, the pair can display a leader-follower relation (LFR) with trajectories of almost synchronized form. However, with larger $d$, although the same LFR is still maintained, the originally similar trajectories turn into different forms. Detailed analysis of their trajectories suggests that the pair might be using U-turns (UTs) to exchange information and to maintain a LFR at the same time. A simulation model based on UTs with inferred and proposed rules is able to reproduce salient features of observed trajectories; indicating that the transition of trajectories can be understood as the result of a change in information exchange between the fish as $d$ increases. Our finding that UTs as important visual signals is consistent with the fact that UTs can induce a large amount of firings in retinas of observing fish.

\end{abstract}
\maketitle
\newpage

%\section{Introduction}
Emergence of complex phenomena \cite {vicsek2012collective} from moving agents has fascinated physicists for a long time. The goal is to understand the observed complexity and the universality with simple rules. This approach has been applied to flocks of birds \cite{mora2016local} and shoals of fish \cite{katz2011inferring}. It was known that three phenomenological rules \cite{reynolds1987flocks} are essential: cohesion, repulsion and alignment. However, in  reality, sensory detection and decision making are the fundamental causes of their behaviors and therefore the emerging complexity.  In other words, information exchange between animals and their processing \cite{couzin2003self, ito2024selective} is at the heart of the problem. Presumably, these processes have been implicitly embedded into the three rules but with no explicit elucidations. It will deepen our understanding of these collective dynamics if one can understand their behaviors in the context of these processes.

Zebrafish frequently perform U-turns (UTs) for various reasons, including escape, exploration, and environmental navigation \cite{lecheval2018social}. While these behaviors are well documented, the underlying dynamics of UTs remain poorly understood. Recently, it is reported that member fish in a group detect and perform UTs\cite{lecheval2017domino} during their group coordination and leadership maintenance in their collective motion. Presumably, these UTs are not just responses of the fish to their environment but might also serve as signals to other fish.  A systematic study of these UTs when the fish are interacting in well controlled environment will offer a window to understand their dynamics in the context of information exchange and decision making.

%When these UTs are observed in a group with close proximity of other fish, it is difficult to single out the causal relation between the UTs performed by a pair of fish because information can come from multiple sources at various distance which is known to affect interaction between two fish \cite{parrish1999complexity}.
%However, very little is known about how information is being exchanged between the animals and how this  information is responsible for the observed behavior \cite{rieucau2015towards}.

Here we report the results of our experiments designed to probe the information exchange between a pair of visually coupled by physically separated zebrafish at different minimal distance ($d$).  We choose to control $d$ because both the repulsion and cohesion rules mentioned above are related to the distance between animals. In fact, we find that as $d$ increases, there is a transition in the forms of the trajectories of the fish pair while the same leader and follower relation (LFR) is still maintained. We will show below that the fish are using UTs to exchange information and to maintain LFR. Our discovery that UTs are important signals is supported by the fact that UTs can induce strong signals in the visual sensory system of the observing fish \cite{schwartz2007synchronized}. We are also able to construct a simulation model based on UTs to reproduce salient features observed in experiments. Our model is an example of using sensory detection and decision makings to describe observed animal behaviors. %This approach not only advances our understanding of zebrafish social interactions but also contributes to broader insights into the physical principles governing animal communication.

%\section{Methods}

\begin{figure}[h!]
	\begin{center}
		\includegraphics[scale=0.4]{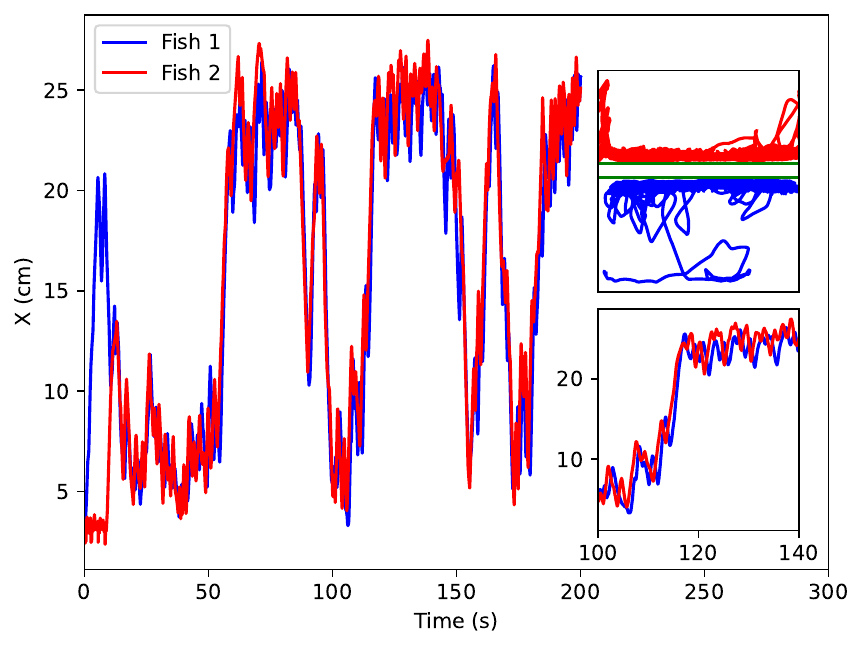}
	\end{center}
	\caption{Time course $X_1(t)$ and $X_2(t)$ in an experiment with $d = 0.4$ cm. Upper inset: trajectories of the two fish in the two channels (not correct aspect ratio); the green lines are the two windows which is $d$ apart. Lower inset: enlarged view of $X_1(t)$ and $X_2(t)$ showing fluctuations near the end of the channel.} 
	\label{Expt_d0}
\end{figure}

%\subsection{Experiment Setup}
Our experiments were carried out in a rectangular ($ 23 cm \times 15 cm$) tank with three channels formed by two parallel windows (0.2 cm thick) $d$ apart and each zebrafish from a pair is placed on each of the lateral channels. The two fish can see each other but they are kept at a minimum distance of $d$ (upper inset of Figure~\ref{Expt_d0}). The water levels in all the three channels are 2 cm. Motions of the two fish are recorded by a CCD camera at 30 frames per second with a spatial resolution of 0.21 mm/pixel. To start an experiment, a fish pair is obtained randomly from a tank with 50 fish and the pair will be returned to the tank after experiment. We find that a settlement time of 0, 2.5, 5 and 10 minutes before the start of an experiment recording did not affect the results. A total of 41 experiments with 7 pairs of fish have been performed with successful trajectories extraction by idtrackerai \cite{van2014scikit}. For the health of the fish, the total experiment period per day is limited to 60 minutes. All experiments were performed at $25 ^\circ C$. Since we cannot compare or average results obtained from different fish pairs, results reported below are based on a single series of experiments performed with the same pair of fish in the same day to avoid variability. However, the  results reported below are the general features common to at least 4 pairs of fish.

%\section{Results}
%\subsection{Three types of interaction}
When the fish pair interact through the transparent windows, their trajectories can be classified as EE, EL and LL. In the EE type both fish are always moving close the barrier trying reach for the other fish (upper inset of Figure~\ref{Expt_d0}). As they are swimming mostly along the length of the barrier (x-axis), their trajectories are dominated by their x-coordinates and denoted as $X_1(t)$ and $X_2(t)$ for Fish 1 and Fish 2 respectively as shown in Figure~\ref{Expt_d0}. Clearly, the dynamics of $X_1$ and $X_2$ are almost synchronized and their trajectories are consisted of fast gliding parts in the middle and oscillating parts close the to ends of the channel (lower inset of Figure~\ref{Expt_d0}). This mutual engaging (EE) interaction is often seen when $d = 0.4 cm$.

\begin{figure}[h!]
	\begin{center}
		\includegraphics[scale=0.4]{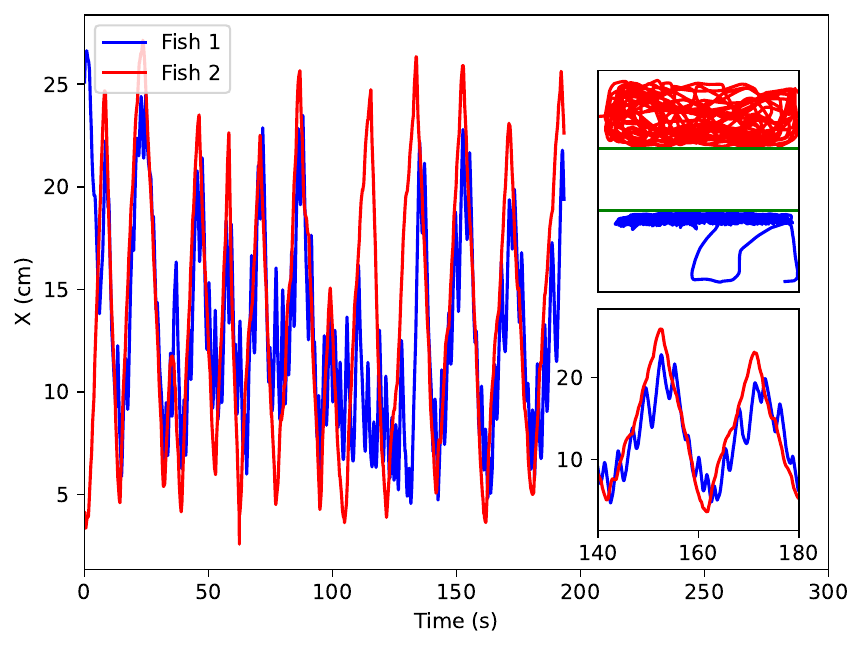}
	\end{center}
	\caption{Time course $X_1(t)$ and $X_2(t)$ in an experiment with $d = 3.4$ cm. Upper inset: trajectories the two fish in the two channels; the green lines are the two windows which is $d$ apart. Lower inset: enlarged view of $X_1(t)$ and $X_2(t)$ showing the difference in dynamics of $X_1$ and $X_2$.} 
	\label{Expt_d3}
\end{figure}

In the EL type (Figure~\ref{Expt_d3}), there is only one engaging fish (E-fish) which always moves close the barrier while the less-engaging fish (L-fish) is not; leading to significance difference in dynamics between $X_1$ and $X_2$. The trajectory (x component) of the L-fish is now mostly gliding along the length of the channel and no oscillatory motions close to the ends of the channel. Interestingly, the E-fish is seen to be following the L-fish but with a faster speed because the E-fish sometimes reverses its direction a couple of times (lower inset of Figure~\ref{Expt_d3}). This EL type is often seen when $d > 0.4$ or $1.4$ cm. The LL type is that both fish are of the less engaging type. More  trajectories of different types can be found in Figure S1, S2  and S3 of the Supplementary Information (SI).

\begin{figure}[h!]
	\begin{center}
 	\includegraphics[scale=0.4]{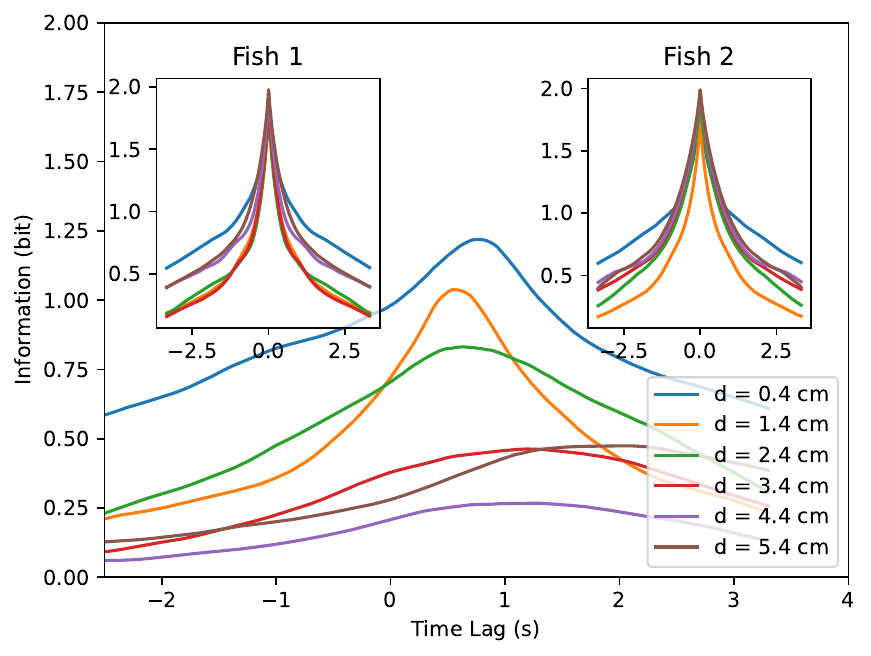}
	\end{center}
	\caption{Effects of minimal separation distance on the cross-TLMI between $X_1(t)$ and $X_2(t)$. In all the TLMI measurements here and below, the noise level of our data are estimated by the TLMI obtained by random shuffling of data and we find that the noise level never exceeds 2\% of the peak of the corresponding TLMI. Both $X_1$ and $X2$ are discretized with a 2 bit resolution. Insets show the auto-TLMI of $X_1$ and $X_2$; indicating that they have similar dynamics at long time scales. All the trajectories are EL type except for $d = 0.4$ and $1.4$ cm which are EE type. }
	\label{Exp_TLMI}
\end{figure}

%\subsection{Time lag mutual information}
Intuitively, the change in information exchange between the fish causes the transition from EE to EL type when $d$ is increased. In order to find out how information is being shared between the two fish as a function of $d$, we have computed time lag mutual information (cross-TLMI) \cite{chou2021anticipation} between $X_1$ and $X_2$ as shown in Figure ~\ref{Exp_TLMI}. From the figure, we find that one fish is always the leader because the peak positions of the cross-TLMI are always located at positive time lags and the leader fish is the L-fish (Fish 2). This observation is supported by the lower insets in Figure~\ref{Expt_d0} and \ref{Expt_d3} in which one can see clearly that the trajectories of Fish 2 is leading those from Fish 1. The peak heights of the cross-TLMI in Figure~\ref{Exp_TLMI} indicate further that their shared information decreases with increasing $d$. Thus, the E-fish is following the L-fish by receiving position information of the L-fish but this information decreases as $d$ increases. 

\begin{figure}[h!]
	\begin{center}
		\includegraphics[scale=0.4]{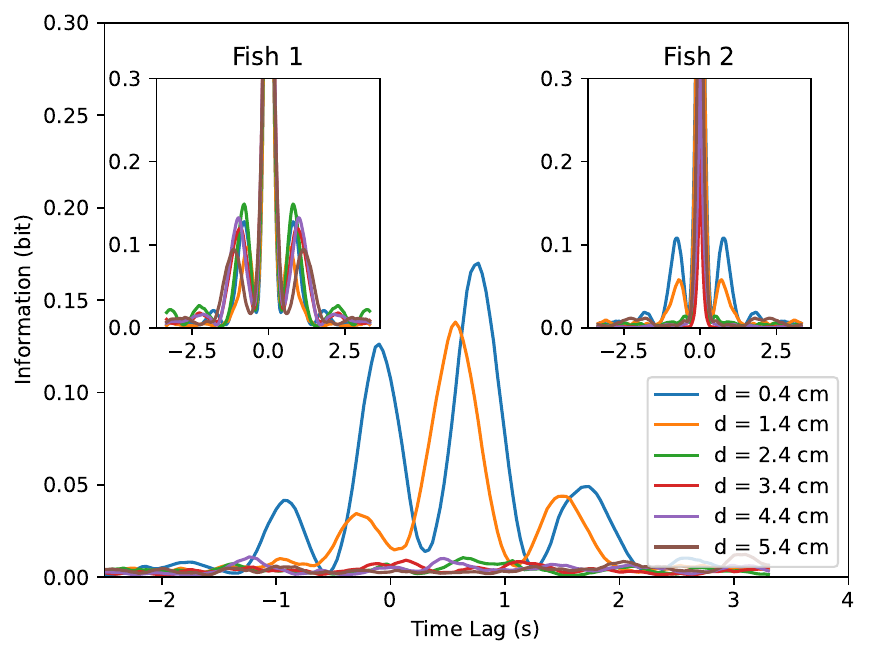}
	\end{center}
	\caption{Effects of detrending on the form of cross-TLMI between $x_1(t)$ and  $x_2(t)$. Insets show the auto-TLMI. The data used are the same as those shown in Figure~\ref{Exp_TLMI} but with the data being detrended with $t_s = 1 s$. A shorter $t_s$ will give smaller cross-TLMI peaks.}
	\label{Exp_dTLMI}
\end{figure}

%\subsection{U-turn detection by Detrending}
The TLMIs reported above are dominated by the large fluctuations in $X_1(t)$ and $X_2(t)$ and are not sensitive to the relatively small fluctuations (UTs) around their mean trajectories. We use detrending to expose these fluctuations. The detrended version of $X_1(t)$ is defined as: $x_1(t) \equiv X_1(t) - S(X_1(t),t_s)$ where $S(X_1(t),t_s)$ is the smoothed version $X_1(t)$ with a smoothing window of size $t_s$ and similarly for $x_2(t)$. Obviously, $x_1(t)$ will contain mainly the fast fluctuations or the UTs shown in the insets of Figure~\ref{Expt_d0} and \ref{Expt_d3}.

Figure~\ref{Exp_dTLMI} shows the cross-TLMI obtained between $x_1(t)$ and $x_2(t)$. There are now side peaks on both side of the central peak of their auto-TLMI; indicating that there are quasi-periodic fluctuations (UTs) in both $x_1(t)$ and $x_2(t)$. For Fish 1, the side peaks appear for all the $d$ while there are side peaks only for $d = 0.4$ and $1.4$ cm for Fish 2.  If these side peaks come from signals used by the fish to send information to the other fish, our result indicates that Fish 1 is sending signal for all $d$ while Fish 2 is doing the same only for $d = 0.4$ and $1.4$ cm. This interpretation is also supported by the fact that Fish 2 are always engaging (Figure S1) for various $d$. With this assumption, it does not come as a surprise that only for $d = 0.4$ and $1.4$ cm, the cross-TLMI between $x_1(t)$ and $x_2(t)$ have significant peaks. That is: the two fish is sharing information only when $d$ is small. Note that the side peaks shown in the auto-TLMI are not unique to this set of data. One can see these side peaks always from the engaging fish in other experiments.

%Similar to the finding in their non-detrended version, the cross-TLMI of $x_1(t)$ and $x_2(t)$ also indicated that the Fish 2 is the leading fish. That is: the information embedded in both positions and UTs of Fish 2 are ahead of those from Fish 1.

%\section{Hypothesis based on observations}

The picture emerges from discussions above is that the transition in the form of the trajectories of the two fish are most likely due the changes in information sharing between them as $d$ is varied because the leader only sends out signals for small $d$. Since UTs are not needed just for following, most likely the UTs of the follower shown in the lower inset of Figure~\ref{Expt_d3} are also used as signals to acknowledge its detection of the leader. This interpretation is consistent with the observation that the leader starts to perform UTs to slow down and reverse its direction close to the boundaries because the leader will then start to see the follower as shown in Figure~\ref{Expt_d0} for small $d$. Once this happens, the two fish will engage in mutual UTs responses seen in Figure~\ref{Expt_d0}. Here, $d$ is important because we assume that the leader will ignore the follower when $d$ is large. 

A phenomenological simulation model based on sensory detection and decision making mechanism is developed to understand our data. In this model, the follower (Fish 1) and leader (Fish 2) move always with constant speeds $v_{1}$ and $v_{2}$ respectively within the interval [0,1]. The update rule is: $X_i(n+1) = X_i(n) + \lambda_i(n) v_i$ ($i = 1$ or $2$) where $X_i(n)$ is the position of the Fish $i$ and $\lambda(n)$ is the direction indicator which can be $\pm 1$ at the $n$-th step. When they reach the boundaries, $\lambda_i(n+1) = - \lambda_i(n)$ to reverse their direction. 

The UTs are implemented as coordinated changes of $\lambda_i(n)$ at desired $n$. One important mechanism in our model is the implementation of a 2UTs; two consecutive U-turns with a total duration of $\tau_{2U}$. These 2UTs are inspired by the quasi-periodic nature of the TLM shown in Figure~\ref{Exp_dTLMI} and are capable of slowing down the fish while still maintaining their motion direction. 

To simulate the leader's behavior observed in experiments, we introduce spontaneous 2UTs of the leader with a probability of $P_s$ per simulation step when the leader is within the slow-down zone ($d_{s}$) near the boundaries (Figure~\ref{Expt_d0}).  Presumably, when the leader is close to the boundaries, it needs to slow down and turn around. Once this happens, the leader will see the follower and then uses these 2UTs to signal its detection of the follower. If the follower is close enough to the leader, it will respond to these 2UTs with its own 2UTs after a time delay of $\tau_d$. These spontaneous 2UTs are absent for the leader when $d$ is large because the follower is too far away.

To maintain the LFR, the follower performs 2UTs to slow down when needed, ensuring a minimal separation of $d_{m}$ from the leader. In summary, the follower responds to the position and UTs of the leader, while the leader moves mostly independently. Details of the parameters and simulation model can be found in the Supplementary Information (SI).

Figure~\ref{Sim_small_d} shows the typical simulated trajectories for the case of small $d$. It can be seen that the trajectories for both fish are almost synchronized with a LFR. Although $v_1$ and $v_2 $ are equal in Figure~\ref{Sim_small_d}, a LFR can still be observed because the follower will execute 2UTs to wait for the leader to lead. These waiting UTs can be seen clearly at time zero in Figure~\ref{Sim_small_d}. Similar to the experimental observations for small $d$, the simulated trajectories are consisted of two parts; linear part in the middle of the [0,1] and fluctuations close to two boundaries of the interval [0,1]. These fluctuations are the spontaneous 2UTs performed by the the leader because $P_s$ is set to $0.15$ here and the follower is also responding with its own 2UTs with a time delay of $\tau_d = 13$ steps. The engagement of these mutual UTs stops when the leader moves out of the slow-down zone where $P_s = 0$.

The cross-TLMI of the detrended trajectories obtained from the simulation is also shown in the lower inset of Figure~\ref{Sim_small_d}. Similar to that from the experiment (Figure~\ref{Exp_dTLMI}),  the cross-TLMI has multiple peaks because of the quasi-periodic nature of the correlated 2UTs from the leader and follower. The location of the main (tallest) peak is determined by the delay in the response of the follower to the UTs of the leader.

\begin{figure}[h!]
	\begin{center}
		\includegraphics[scale=0.4]{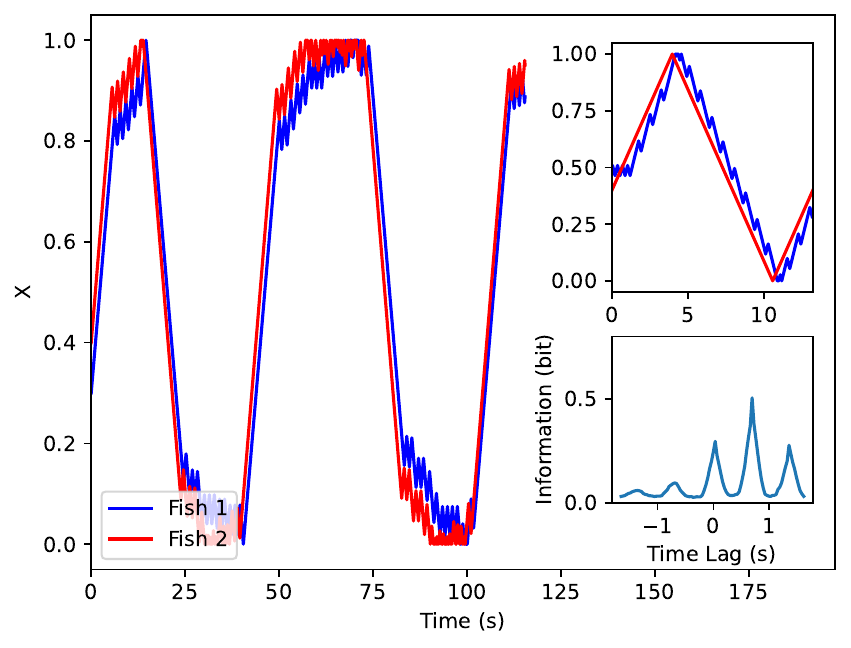}
	\end{center}
	\caption{Simulated trajectories of the follower (Fish 1) and leader (Fish 2) in the interval [0,1] for the case of small $d$. Parameters are: $v_1 = v_2 = 0.003$, $d_s = 0.1$, $P_s$, $\tau_{2U} = 40$, $\tau_d = 13$ and $d_m = 0.05$ Lower inset: the cross-TLMI for this small $d$ case and the time delay of tallest peak is determined by $\tau_d$ between the UTs of the $\alpha$ and follower. Upper inset: simulated trajectories for the case of  large $d$. Parameters  are the same as stated above except for $v_{2} = 0.005$ and $v_{1} = 0.009$ and $P_{s} = 0$. In the graphs, one simulation step is set to $0.033 s$.}
	\label{Sim_small_d}
\end{figure}

Simulated trajectories for the case of large $d$ are shown in the upper inset of Figure~\ref{Sim_small_d}. Here, the leader is not engaging with the follower and it will not perform any spontaneous 2UTs. In this case, $P_s = 0$ and therefore the leader will be moving with a linear trajectory. As for the follower, we need to set a higher speed for it because it will then perform 2UTs to implement the LFR by keeping a minimal distance $d_m$ from the leader as observed in the experiment. There would have been more 2UTs shown in the inset if the difference in $v_1$ and $v_2$ were larger. Figure~\ref{Sim_small_d} reproduces the essential features of the observed trajectories shown in Figure~\ref{Expt_d0}, \ref{Expt_d3} and \ref{Exp_dTLMI}. 

From the discussions above, it is clear that the apparent complex transition of the trajectories of fish pair when $d$ is changed can be understood from our simple rules. Our model is an example of using sensory detection and decision makings (rules) to understand emerging behavior of animals \cite{ito2024selective}; without ad hoc forces. While our model captures many aspects of these behaviors, it does not reproduce all the features. For instance, the form of the trajectories of the follower fish between $100$ to $140 s$ in Figure~\ref{Expt_d3} are not reproduced. This suggests that interactions between animals are intricate, and our simple rules likely apply only within specific experimental conditions. We do not model the response of the leader to the UTs of follower because the leader cannot see the follower if they are swimming in the same direction. When they are both performing UTs close to the boundaries as shown in Figure~\ref{Expt_d0}, they should be able to see each other. Interestingly, this is the case where anticipatory dynamics \cite{chen2023detection} could have taken place but difficult to detect.

Key insights emerge when re-examining the phenomenological cohesion and repulsion rules. The parameter $d_m$ in our model does not only enforce the repulsion rule but also provide long enough distance between the animals such that the observing fish can detect the overall movement of the leading fish.  If $d_m$ is too small, the observing fish will be seeing parts of the  body of the leading fish whose movement might be too fast to decode. As for the cohesion rule,  it is enforced by the maintenance of the LFR by the follower only.  In our experiments, it seems that LFR  is determined by intrinsic properties of the fish pair. In our experiments, we found that some pairs just do not engage to display LFR as in the LL type trajectories (see Figure S3 in SI).  It would be interesting if one can also deduce that the followers are the ones to maintain LFR in a school of fish.  %Since we are working in a 1D model, the alignment rule is not needed. 
 
Of all the rules in our model, the most intriguing one is the 2UTs inspired by the quasi-periodic nature of the motion of the engaging fish. This pattern likely arises when the follower swims next to a transparent wall while attempting to reach the leader.  Since the fish can only communicate visually, our assumption that these UTs are signals is consistent with the fact that motion reversal will trigger a large amount of firing in retina  \cite{schwartz2007synchronized} of the observing fish. Therefore, these 2UTs can serve both as a speed controller and a signal generator. Recently, it is believed that emerging collective behavior might have neurological origin \cite{yu2024understanding}. It would be extremely interesting if these 2UTs, which have a characteristic time of about 1 s, also have neural basis \cite{portugues2009neural}. 

%Finally, we must stress that the situation in our experiment is quite different from the normal case when the fish are swimming and interacting freely. The UTs modeled here might not be the same the UTs \cite{lecheval2017domino} seen in the coordination of freely moving fish groups.

%The maintenance of leader-follower relationships despite changes in trajectory morphology is also noteworthy, as it contributes to our understanding of the dynamics of these relationships. Ultimately, this research has far-reaching implications, with possible connections to human social behavior and psychology. By elucidating the mechanisms underlying social interaction and communication in zebrafish, this study has paved the way for future investigations into the complexities of animal sociality, and may inspire new avenues of research into the intricacies of human social behavior. Overall, this work demonstrates the value of interdisciplinary research in advancing our understanding of complex behavioral phenomena.

This work has been supported by the MOST of ROC under the grant number 111-2112-M-001-032.

%\newpage
%\bibliography{SBRef.bib}
\bibliographystyle{prsty}
\bibliography{fish_ref01}

\begin{thebibliography}{10}

\bibitem{vicsek2012collective}
T. Vicsek and A. Zafeiris, Physics reports {\bf 517},  71  (2012).

\bibitem{mora2016local}
T. Mora {\it et~al.}, Nature physics {\bf 12},  1153  (2016).

\bibitem{katz2011inferring}
Y. Katz {\it et~al.}, Proceedings of the National Academy of Sciences {\bf
  108},  18720  (2011).

\bibitem{reynolds1987flocks}
C.~W. Reynolds,  in {\em Proceedings of the 14th annual conference on Computer
  graphics and interactive techniques} (PUBLISHER, ADDRESS, 1987), pp.\ 25--34.

\bibitem{couzin2003self}
I.~D. Couzin {\it et~al.}, Advances in the Study of Behavior {\bf 32},  10
  (2003).

\bibitem{ito2024selective}
S. Ito and N. Uchida, PNAS nexus {\bf 3},  264  (2024).

\bibitem{lecheval2018social}
V. Lecheval {\it et~al.}, Proceedings of the Royal Society B: Biological
  Sciences {\bf 285},  20180251  (2018).

\bibitem{lecheval2017domino}
V. Lecheval {\it et~al.}, BioRxiv  138628  (2017).

\bibitem{schwartz2007synchronized}
G. Schwartz {\it et~al.}, Neuron {\bf 55},  958  (2007).

\bibitem{van2014scikit}
S. Van~der Walt {\it et~al.}, PeerJ {\bf 2},  e453  (2014).

\bibitem{chou2021anticipation}
P.-Y. Chou {\it et~al.}, Physical Review E {\bf 103},  L020401  (2021).

\bibitem{chen2023detection}
W.-J. Chen {\it et~al.}, Entropy {\bf 26},    (2023).

\bibitem{yu2024understanding}
J.-H. Yu, J.~L. Napoli, and M. Lovett-Barron, Current Opinion in Neurobiology
  {\bf 86},  102866  (2024).

\bibitem{portugues2009neural}
R. Portugues and F. Engert, Current opinion in neurobiology {\bf 19},  644
  (2009).

\end{thebibliography}

\end{document}